\documentstyle[12pt]{article}

\author{Chi Au and Bao-Heng Zhao\thanks{Permanent address: Department of
Physics, Graduate School, Chinese Academy of Sciences, P. O. Box 3908, Beijing
100039, China}\\
Department of Applied Mathematics\\
Hong Kong Polytechnic University, Hong Kong}
\title{From ODLRO to the Meissner Effect and Flux Quantization
}

\begin{document}

\maketitle
\begin{abstract}
It has been shown that the electron system with ODLRO in the reduced density
matrix $\rho _2$ can not support a uniform magnetic field, i.e., ODLRO\ in $%
\rho _2$ implies the Meissner effect \cite{2}\cite{3}, furthermore, the
magnetic field trapped in the system is quantized\cite{3}. This note extends
above results in two cases. We show that (1) the system with ODLRO in $\rho
_2$ can not support a non uniform, cylindrically symmetric magnetic field;
(2) the system with ODLRO in $\rho _2$ can not support a magnetic field
slowly varying in space, and the magnetic flux trapped in it is quantized.

\end{abstract}

\newpage\

\begin{center}
{\bf Introduction}
\end{center}

In 1962 C. N. Yang proposed that the superconductivity is characterized by
the existence of off-diagonal long-range order (ODLRO) in the two-particle
reduced density matrix $\rho _2$ for a fermion system\cite{1}. However,
there was no clear-cut proof to show directly that ODLRO in $\rho _2$ leads
to superconductivity for a long time. In recent works \cite{2}\cite{3}, it
has been proved that ODLRO in $\rho _2$ implies that a non vanishing{\it \ }%
uniform magnetic field ${\bf B}$ can not exist in an electron system, i.e.,
there is the Meissner effect. Furthermore, it has been shown that the
magnetic flux trapped in the system is quantized\cite{3}. However, the
question of whether a non uniform magnetic field can be expelled from the
system with ODLRO in $\rho _2$ is still open. The present work extends \cite
{2} and \cite{3} in two cases. We show that 1. the system with ODLRO in $%
\rho _2$ can not support a non uniform, cylindrically symmetric magnetic
field; 2. the system with ODLRO in $\rho _2$ can not support a non uniform
magnetic field, which is slowly varying in space, and the magnetic flux
trapped in the system is quantized.

\begin{center}
{\bf System with Cylindric Symmetry}
\end{center}

Consider the case that $N$ electrons are in a circular cylinder. We chose a
cylindric coordinates with z-axis along the central line of the cylinder. In
which ${\bf r}=(R,\theta ,z),$ and ${\bf \hat R}${\bf $,\hat \theta $ }and $%
{\bf \hat z}$ are unit vectors along the $R,\theta $ and $z$ direction,
respectively. Suppose that the magnetic field is cylindrically symmetric,
and the gauge to be \cite{4}
\begin{equation}
\label{1}{\bf A(r)=-}\theta RB(R){\bf \hat R},
\end{equation}
where $B(R)$ is a continuous function. From (1), we have ${\bf B(r)=\nabla
\times A(r)}=B(R){\bf \hat z.}$ Suppose that the system undergoes an
infinitesimal rotation ${\bf \hat z}\delta \theta ,$
\begin{equation}
\label{2}
\begin{array}{c}
{\bf A(r)\rightarrow -(}\theta -\delta \theta )RB(R){\bf \hat R}={\bf %
A(r)+\nabla }\chi _{\delta \theta }({\bf r}), \\ \chi _{\delta \theta }({\bf %
r})=\delta \theta \int\limits_{R_0}^RR^{\prime }B(R^{\prime })dR^{\prime },
\end{array}
\end{equation}
where $R_0$ is a constant. Eq.(2) shows that the rotation induces a gauge
transformation. We shall see that it enable us to show that ${\bf B}({\bf r}%
)=0$ as the system has ODLRO in the two particle reduced density matrix $%
\rho _2.$

The Schr\"odinger equation of the electrons is
\begin{equation}
\label{3}
\begin{array}{c}
\frac 1{2m}\sum\limits_j[\frac \hbar i
{\bf \nabla }_j+\frac ec{\bf A(r}_j)]^2\psi _n({\bf r}_1,...,{\bf r}_N)+U(%
{\bf r}_1,...,{\bf r}_N)\psi _n({\bf r}_1,...,{\bf r}_N) \\ =E_n\psi _n({\bf %
r}_1,...,{\bf r}_N).
\end{array}
\end{equation}
We assume that the potential $U({\bf r}_1,...,{\bf r}_N)$ depends only on
the distances between electrons. Under the rotation, (3) becomes
\begin{equation}
\label{4}
\begin{array}{c}
\frac 1{2m}\sum\limits_j[\frac \hbar i
{\bf \nabla }_j+\frac ec{\bf A(r}_j)]^2\psi _n^{\prime }({\bf r}_1,...,{\bf r%
}_N) \\ +U({\bf r}_1,...,{\bf r}_N)\psi _n^{\prime }({\bf r}_1,...,{\bf r}%
_N)=E_n\psi _n^{\prime }({\bf r}_1,...,{\bf r}_N),
\end{array}
\end{equation}
where
\begin{equation}
\label{5}
\begin{array}{c}
\psi _n^{\prime }(
{\bf r}_1,...,{\bf r}_N)=\psi _n({\bf r}_1-\delta {\bf r}_1,...,{\bf r}%
_N-\delta {\bf r}_N)\exp [\frac{ie}{c\hbar }\sum\limits_j\chi _{\delta
\theta }({\bf r}_j)],\, \\ \,\delta {\bf r_i=\hat \theta }_iR_i\delta \theta
{}.
\end{array}
\end{equation}
Both \{ $\psi _n$\} and \{$\psi _n^{\prime }$\} are complete sets of
orthonormal eigenfunctions of the Hamiltonian, which are single-valued in $%
{\bf r}_i.$

The element of the reduced density matrix for two particles can be expressed
as
$$
\rho _2({\bf r^{\prime }}_1,{\bf r^{\prime }}_2;{\bf r}_1,{\bf r}_2)=\int
\cdots \int \frac{d{\bf r}_3\cdots d{\bf r}_N}{(N-2)!}\frac
1Z\sum_nexp(-E_n/kT)
$$
\begin{equation}
\label{6}\times \psi _n({\bf r^{\prime }}_1,{\bf r}_2^{\prime },{\bf r}%
_3,\cdots ,{\bf r}_N)\psi _n^{*}({\bf r}_1,{\bf r}_2,{\bf r}_3,\cdots ,{\bf r%
}_N),
\end{equation}
where {\it Z} is the partition function, and {\it T} the temperature.
Substituting $\psi _n^{\prime }$ for $\psi _n$ makes no change for $\rho _2,$
since $\rho _2$ is basis independent. So we also have
$$
\rho _2({\bf r}_1^{\prime },{\bf r}_2^{\prime };{\bf r}_1,{\bf r}_2)=\int
\cdots \int \frac{d{\bf r}_3\cdots d{\bf r}_N}{(N-2)!}\frac
1Z\sum_nexp(-E_n/kT)
$$
\begin{equation}
\label{7}\times \psi _n^{\prime }({\bf r^{\prime }}_1,{\bf r^{\prime }}_2,%
{\bf r}_3,\cdots ,{\bf r}_N)\psi _n^{\prime }{}^{*}({\bf r}_1,{\bf r}_2,{\bf %
r}_3,\cdots ,{\bf r}_N),
\end{equation}
Expressing the primed $\psi _n^{\prime }$ in (7) in terms of the non primed $%
\psi _n$, we obtain
$$
\rho _2({\bf r^{\prime }}_1,{\bf r^{\prime }}_2;{\bf r}_1,{\bf r}_2)=exp\{
\frac{ie}{c\hbar }[\chi _{\delta \theta }({\bf r^{\prime }}_1)+\chi _{\delta
\theta }({\bf r^{\prime }}_2)-\chi _{\delta \theta }({\bf r}_1)-\chi
_{\delta \theta }({\bf r}_2)]\}
$$
\begin{equation}
\label{8}\times \int \cdots \int \frac{d{\bf r}_3\cdots d{\bf r}_N}{(N-2)!}%
\frac 1Z\sum_nexp(-E_n/kT)
\end{equation}
$$
\begin{array}{c}
\times \psi _n(
{\bf r^{\prime }}_1-\delta {\bf r^{\prime }}_1,{\bf r^{\prime }}_2-\delta
{\bf r^{\prime }}_2,{\bf r}_3-\delta {\bf r}_3,\cdots ,{\bf r}_N-\delta {\bf %
r}_N) \\ \times \psi _n^{*}({\bf r}_1-\delta {\bf r}_1,{\bf r}_2-\delta {\bf %
r}_2,{\bf r}_3-\delta {\bf r}_3,\cdots ,{\bf r}_N-\delta {\bf r}_N).
\end{array}
$$
Shift the integration variables, $\theta _i$ $\rightarrow \theta _i+\delta
\theta ,\,i=3,...,N.$ Note that ${\bf r}_i-\delta {\bf r}_i=(R_i,\theta
_i-\delta \theta ,z_i)$. We obtain a relation for $\rho _2$ at different
space points:
\begin{equation}
\label{9}
\begin{array}{c}
\rho _2(
{\bf r^{\prime }}_1,{\bf r^{\prime }}_2;{\bf r}_1,{\bf r}_2)=exp\{\frac{ie}{%
c\hbar }[\chi _{\delta \theta }({\bf r^{\prime }}_1)+\chi _{\delta \theta }(%
{\bf r^{\prime }}_2)-\chi _{\delta \theta }({\bf r}_1)-\chi _{\delta \theta
}({\bf r}_2)]\} \\ \times \rho _2({\bf r}_1^{\prime }-\delta {\bf r}%
_1^{^{\prime }},{\bf r}_2^{\prime }-\delta {\bf r}_2^{^{\prime }};{\bf r}%
_1-\delta {\bf r}_1,{\bf r}_2-\delta {\bf r}_2).
\end{array}
\end{equation}

By definition, the ODLRO in $\rho _2$ means that
\begin{equation}
\label{10}\rho _2({\bf r}_1^{\prime },{\bf r}_2^{\prime };{\bf r}_1,{\bf r}%
_2)\rightarrow \Phi ({\bf r^{\prime }}_1,{\bf r^{\prime }}_2)\Phi ^{*}({\bf r%
}_1,{\bf r}_2),
\end{equation}
as $\left| {\bf r^{\prime }}_i-{\bf r}_j\right| ,i,$ $j=1,2,$ approaches
macroscopic distances, and $\Phi ({\bf r}_1,{\bf r}_2)$ does not vanish as $%
\left| {\bf r}_1-{\bf r}_2\right| $ and $\left| {\bf r^{\prime }}_1-{\bf %
r^{\prime }}_2\right| $ remain finite in microscopic scale. Where $\Phi $ is
the eigenfunction of $\rho _2$ with the largest eigenvalue. If we assume
ODLRO in $\rho _2,$ (8) and (9) lead to
\begin{equation}
\label{11}
\begin{array}{c}
\Phi (
{\bf r}_1,{\bf r}_2)=\exp [i\alpha (\delta \theta )]\exp \{\frac{ie}{c\hbar }%
[\chi _{\delta \theta }({\bf r}_1)+\chi _{\delta \theta }({\bf r}_2)]\} \\
\times \Phi ({\bf r}_1-\delta {\bf r}_1,{\bf r}_2-\delta {\bf r}_2),
\end{array}
\end{equation}
where $\alpha $$(\delta \theta )$ is a real function of $\delta \theta ,$
and it is independent of ${\bf r}_1,{\bf r}_2.$ It is easily seen that $%
\alpha $$(\delta \theta +\delta \theta ^{\prime })=\alpha $$(\delta \theta
)+\alpha $$(\delta \theta ^{\prime }),$ $\alpha (0)=0.$

Rotating the system successively, as $\sum \delta \theta =2\pi $, we have $%
\sum \delta {\bf r}_1=\sum \delta {\bf r}_2=0.$ $\Phi ({\bf r}_1,{\bf r}_2)$
is single-valued because $\psi _n$ is single-valued. Thus, we obtain from
(10)
\begin{equation}
\label{12}\exp [i\alpha (2\pi )]\exp \{\frac{ie}{c\hbar }2\pi
[\int\limits_{R_0}^{R_1}RB(R)dR+\int\limits_{R_0}^{R_2}RB(R)dR]\}=1.
\end{equation}
(12) leads to
\begin{equation}
\label{13}\frac{c\hbar }e\alpha (2\pi )+2\pi
[\int\limits_{R_0}^{R_1}RB(R)dR+\int\limits_{R_0}^{R_2}RB(R)dR]=n\frac{ch}%
e,\,\,\,n=integers.
\end{equation}

On the left hand side of (13), $R_1$ and $R_2$ can be changed
continuously. Let $R_1\rightarrow R_1+dR_1,$ the left hand of (13) gets an
infinitesimal increment $2\pi R_1B(R_1)dR_1$. On the other hand, the right
hand side only takes discrete values. To be consistent, we have to take $%
B(R_1)=0.$ Since $R_1$ is arbitrary in the system, we obtain
\begin{equation}
\label{14}{\bf B(r})=0.
\end{equation}
Thus a non vanishing cylindrically symmetric magnetic field can not exist in
the system with ODLRO in $\rho _2.$ In other words, the magnetic field is
expelled from the system, i.e., there is the Meissner effect.

\begin{center}
{\bf Electrons in the Magnetic Field Slowly Varying in Space}
\end{center}

Consider the case of electrons in a magnetic field which is slowly varying
in space. The vector potential can be written as
\begin{equation}
\label{15}{\bf A}({\bf r})={\bf A}_0({\bf r})+{\bf \nabla }\varphi ({\bf r}%
),
\end{equation}
and ${\bf B(r})={\bf \nabla }\times {\bf A}_0({\bf r}).$ In general, $%
\varphi ({\bf r})$ is multi-valued, however, ${\bf \nabla }\varphi ({\bf r})$
is single-valued.

Under an infinitesimal translation $\delta {\bf r}$,

\begin{equation}
\label{16}{\bf A(r)}\rightarrow {\bf A}_0({\bf r}-\delta {\bf r})+\nabla
\varphi ({\bf r}-\delta {\bf r})={\bf A(r)}-\delta {\bf r}\cdot {\bf \nabla }%
[{\bf A}_0({\bf r})+{\bf \nabla }\varphi ({\bf r})].
\end{equation}
In (16)
\begin{equation}
\label{17}
\begin{array}{c}
\delta
{\bf r}\cdot \nabla {\bf A}_0=\nabla (\delta {\bf r\cdot A}_0)-\delta {\bf %
r\times B} \\ =\nabla (\delta {\bf r\cdot A}_0)-\nabla [\delta {\bf r\cdot
(B\times r)]}+[(\delta {\bf r}\cdot \nabla ){\bf B]\times r\ +\ }\delta {\bf %
r}\times [({\bf r}\cdot \nabla ){\bf B].}
\end{array}
\end{equation}
Provided that ${\bf B(r})$ is slowly varying, such that
\begin{equation}
\label{18}\left| {\bf \delta r\times B}\right| >>\left| [(\delta {\bf r}%
\cdot \nabla ){\bf B]\times r}+\delta {\bf r}\times [({\bf r}\cdot \nabla )%
{\bf B]}\right|
\end{equation}
i.e., roughly speaking, $\left| \nabla B_i\right| /\left| {\bf B}\right| ,$ $%
i=x,y,z,$ is much less than $L^{-1},$ where {\it L} is the linear dimension
of the system. Under this condition, we can write
\begin{equation}
\label{19}\delta {\bf r}\cdot \nabla {\bf A}_0=\nabla (\delta {\bf r\cdot A}%
_0)-\nabla [\delta {\bf r\cdot (B\times r)]},
\end{equation}
and (16) becomes
\begin{equation}
\label{20}{\bf A(r)}\rightarrow {\bf A(r)+\nabla [}\delta {\bf r\cdot
(B\times r)-}\delta {\bf r\cdot A}_0-\delta {\bf r}\cdot {\bf \nabla }%
\varphi ({\bf r})].
\end{equation}
We see that the infinitesimal translation induces a gauge transformation.

In the Hamiltonian of the system,
\begin{equation}
\label{21}H=\sum\limits_j\frac 1{2m}[-i\hbar {\bf \nabla }_j+\frac ec{\bf A}(%
{\bf r}_j)]^2+U({\bf r}_1,...,{\bf r}_N),
\end{equation}
we assume that the potential $U({\bf r}_1,...,{\bf r}_N)$ depends only on
the distances between electrons. Suppose that $\{\psi _n\}$ is a complete
set of single-valued orthonormal eigenfunctions of the Hamiltonian. Then $%
\{\psi _n^{\prime }\}$ is also a complete set of single-valued orthonormal
eigenfunctions of the Hamiltonian\cite{3},where
\begin{equation}
\label{22}\psi _n^{\prime }({\bf r}_1,...{\bf ,r}_N)=e^{\frac{ie}{c\hbar }%
\sum_j\chi _{\delta {\bf r}}({\bf r}_j)}\psi _n({\bf r}_1-\delta {\bf r},...,%
{\bf r}_N-\delta {\bf r}),
\end{equation}
and
\begin{equation}
\label{23}\chi _{\delta {\bf r}}({\bf r}_j)=\delta {\bf r\cdot }[{\bf B(r}_j%
{\bf )\times r}_j-{\bf A}_0({\bf r_j})-\nabla _j\varphi ({\bf r}_j)].
\end{equation}

Repeating the discussion similar to that leads to (8) ( see also \cite{3}),
we have
\begin{equation}
\label{24}
\begin{array}{c}
\rho _2(
{\bf r^{\prime }}_1,{\bf r^{\prime }}_2;{\bf r}_1,{\bf r}_2)=exp\{\frac{ie}{%
c\hbar }[\chi _{\delta {\bf r}}({\bf r^{\prime }}_1)+\chi _{\delta {\bf r}}(%
{\bf r^{\prime }}_2)-\chi _{\delta {\bf r}}({\bf r}_1)-\chi _{\delta {\bf r}%
}({\bf r}_2)]\} \\ \times \rho _2({\bf r}_1^{\prime }-\delta {\bf r},{\bf r}%
_2^{\prime }-\delta {\bf r};{\bf r}_1-\delta {\bf r},{\bf r}_2-\delta {\bf r}%
).
\end{array}
\end{equation}
If there is ODLRO in $\rho _2,$ (10) is valid, and
\begin{equation}
\label{25}\Phi ({\bf r}_1,{\bf r}_2)=\exp [i\alpha (\delta {\bf r})]\exp \{
\frac{ie}{c\hbar }[\chi _{\delta {\bf r}}({\bf r}_1)+\chi _{\delta {\bf r}}(%
{\bf r}_2)]\}\Phi ({\bf r}_1-\delta {\bf r},{\bf r}_2-\delta {\bf r}),
\end{equation}
where $\alpha (\delta {\bf r})$ is real, and not dependent on ${\bf r}_1,%
{\bf r}_2.$ It is easily seen that $\alpha (\delta {\bf r}_1+\delta {\bf r}%
_2)=\alpha (\delta {\bf r}_1)+\alpha (\delta {\bf r}_2),$ and $\alpha (0)=0$$%
.$ Repeating successively for infinitesimal displacements along a closed
path, and noting the single-valuedness of $\Phi ({\bf r}_1,{\bf r}_2),$ we
obtain from (25) the relation
\begin{equation}
\label{26}\exp \{i\sum \alpha (\delta {\bf r})\}\exp \{\frac{2ie}{c\hbar }%
\oint d{\bf r}\cdot [{\bf B(r)\times r}-{\bf A}_0({\bf r})-\nabla \varphi (%
{\bf r})]\}=1,
\end{equation}
where the summation $\sum \alpha (\delta {\bf r})$ is over the closed path.
In a simply connected region, around a sufficiently small contour, on which
{\bf B }can be considered as a constant vector, we have $\oint {\bf dr\cdot
[B(r)\times r]}=2\phi ,$ $\oint d{\bf r}\cdot {\bf A}_0({\bf r})=\phi ,$
where $\phi $ is the magnetic flux through the surface spanned by the small
contour, and $\oint d{\bf r}\cdot \nabla \varphi ({\bf r})=0,$ $\sum \alpha
(\delta {\bf r})=0$. We obtain,
\begin{equation}
\label{27}\exp \{\frac{2ie}{c\hbar }\oint d{\bf r}\cdot [{\bf B(r)\times r}-%
{\bf A}_0({\bf r})]\}=\exp \{\frac{2ie}{c\hbar }\phi \}=1.
\end{equation}
Thus
\begin{equation}
\label{28}\phi =n\frac{ch}{2e},\,\,\,n=integers.
\end{equation}

The contour can be deformed continuously. If ${\bf B}\neq 0,$ the flux $\phi
$ can be changed continuously. Which is in contradiction to the right hand
side of (28). To be consistent, we conclude that ${\bf B=}0$. It means that
the system with ODLRO in $\rho _2$ can not support a magnetic field slowly
varying in space, i.e., there is the Meissner effect.

Since ${\bf B}=0,$ we can take ${\bf A}_0=0.$ When the path is not simply
connected, e.g., it winds around an inaccessible tunnel, the path can not be
deformed across this inaccessible region. In this case we still have $\sum
\alpha (\delta {\bf r})=0\cite{5}$, from (26), we obtain $\exp \{\frac{2ie}{%
c\hbar }\oint d{\bf r\cdot }\nabla \varphi ({\bf r})]\}=1,$ thus
\begin{equation}
\label{29}\oint d{\bf r\cdot }\nabla \varphi ({\bf r})=n\frac{ch}{2e}%
,\,\,\,n=integers.
\end{equation}
This equation shows that the magnetic flux trapped in the tunnel is
quantized.

In summary, we have shown that if there is ODLRO in the reduced density
matrix $\rho _2$, 1. the system can not support a cylindrically symmetric
magnetic field; 2. the system can not support a slowly varying magnetic
field. In these two cases, we have derived the Meissner effect from the
existence of ODLRO\ in $\rho _2$. Furthermore, in the latter case, we have
shown that the magnetic flux trapped in the system is quantized. We know
that the magnetic field is not vanishing, e.g., in the boundary region in
superconductor. This fact is not in contradiction to our result, since the
non vanishing magnetic field is in the region where it varies rapidly.

\begin{center}
{\bf Acknowledgment}
\end{center}

This work is supported in part by The Hong Kong Polytechnic University. One
of the authors (B. H. Zhao) is grateful to the hospitality of the Department
of Applied Mathematics, The Hong Kong Polytechnic Universty. He is also
supported in part by NNSF of China.

\end{document}